\magnification=1200

\baselineskip=20pt

\line{\hfil MRI-PHY/P970407}

\def\vphitz{<\phi_2^0\phi_2^0>}
\def\vphitzs{<\phi_2^{0*}\phi_2^{0*}>}

\def\vPoPt{<\Phi_1\Phi_3>}
\def\vPtPt{<\Phi_2\Phi_3>}
\def\absvo{\vert v_1 \vert}
\def\absvt{\vert v_2 \vert}
\def\nbn{\bar {N}N}
\def\nbign{\bar {N}\imath \gamma_5 N}
\def\nbigtn{\bar {N}\imath\gamma_5\tau_3 N}
\def\cz{C^0_{S-PS}}

\def\gf{\gamma_5}

\centerline{\bf P and T odd nucleon-nucleon interaction in}
\centerline{\bf a two higgs doublet model}
\vskip .4truein
\centerline{\bf Uma Mahanta}
\centerline{\bf Mehta Research Institute}
\centerline{\bf Chhatnag Road, Jhusi}
\centerline{\bf Allahabad-221506}
\centerline{\bf India}
\vskip 1truein
\centerline{\bf Abstract}

A two higgs doublet model with tree level natural flavor conservation
can lead to P and T odd nucleon-nucleon (NN) interaction through tree level
Feynman diagrams. In this article we estimate the magnitudes of the parameters
that characterize the strength of the S-PS type of NN interaction. Stringent
constraints on the parameters can be derived from the measurement of EDM's
of Xe, Cs and Hg atoms in the ground state. We show that the predicted
values of the parameters lie close to such constraints provided 
${\absvt\over \absvo }$ is large. 
\vfill
\eject

Although in the SM [1] only one higgs doublet was used to give mass to the
weak gauge bosons and fermions, there is no good theoretical justification
for using only one doublet. In fact some of the attractive features of the
SM like the tree level custodial symmetry and natural flavor conservation [NFC]
are preserved in suitably chosen versions of the two higgs doublet model
 [THDM].
The minimal supersymmetric standard model [2] which has been extremely
successful in explaining all the experimental data so far, requires
two higgs doublets with Y=$\pm 1$ to preserve supersymmetry.       

In this article we shall consider a general THDM in which tree level
natural flavor conservation is guaranteed by requiring that $\phi_1 $ couples
only to RH $I_3=-{1\over 2}$ fermions and $\phi_2$ couples only to RH
$I_3={1\over 2}$ fermions [3]. We shall show that such a model can give rise
to four fermionic S-PS  nucleon-nucleon interaction which are strong
enough to produce observable P and T odd effects in high precision
atomic and molecular experiments. The coupling of the neutral
higgs bosons $\phi_1^0$ and $\phi_2^0$ to the $Q={2\over 3}$ and
$Q=-{1\over 3}$ quarks are given by
$$ L_y=-[{1\over v_1}\bar {D}^i_L m_{iD} D^i_R \phi_1^0
+{1\over v_2}\bar {U}^j_R m_{jU} U^j_L \phi_2^0 +h.c.]. \eqno(1)$$ 
where we have denoted  the $Q={2\over 3}$ and $Q=-{1\over 3}$ quarks  
generically by U and D respectivvely. i and j are the generation
indices. Consider first the contribution of $Q={2\over 3}$ quarks to tthe
P and T odd NN interaction. We have 
$$L_1^{UU}={m_{iU}m_{jU}\over 8v_2^2}\bar{U}_i(1-\gf )U_i
\bar{U}_j(1-\gf )U_j \imath \vphitz . \eqno(2a)$$
and  
$$L_2^{UU}={m_{iU}m_{jU}\over 8v_2^{*2}}\bar{U}_i(1+\gf )U_i
\bar{U}_j(1+\gf )U_j \imath \vphitzs . \eqno(2b)$$

The P and T odd S-PS mixing term $\bar{U}_i U_i
\bar{U}_j\gamma_5 U_j$ will arise from the Im parts of the propagators
$\imath \vphitz $
  and $\imath \vphitzs $. Note that $\imath <\phi_2^0\phi_2^{0*}>$ is real
and therefore it does not contribute to $L^{UU}_{PTV}$.
 In unitary  gauge [4] we have
 $\phi_1^0={v_1\over \sqrt {2}\absvo  }[\Phi_1-\imath 
 {\absvt \over v}\Phi_3]$ and
  $\phi_2^0={v_2\over \sqrt {2}\absvt }[\Phi_2-\imath 
 {\absvo \over v}\Phi_3]$ where $\Phi_1$, $\Phi_2$ and $\Phi_3$
 are real scalar fields. It then follows that 
 
 $$Im {\imath \vphitz\over v_2^2}=-Im
  {\imath \vphitzs\over v_2^{*2}}={\absvo
  \over \absvt ^2 v}\imath<\Phi_2\Phi_3>.\eqno(3)$$
   
 and
 $$L^{UU}_{PTV}=-  {\absvo
  \over 2 \absvt ^2 v}\imath<\Phi_2\Phi_3>
  \bar{U}_im_{iU} U_i\bar{U}_jm_{jU}\imath \gf  U_j .\eqno(4)$$
  
where a summation over generation indices i and j is implied. In order to
convert the above P and T odd interaction between a pair of U quarks into 
P and T odd NN interaction we have to find the matrix elements of
$\sum_i \bar {U}_i m_{iU} U_{i}$ and
 $\sum_j\bar{U}_j m_{jU}\imath \gf U_j$ between nucleon states [5].
 The light quarks u and d, which are the valence quarks for nucleon and 
 whose masses are much small compared to the QCD
 scale $\Lambda_c$, contribute very little to the scalar matrix element.
 The heavy quarks s, c, b and t on the other hand exist inside the nucleon
 only as virtual states and therefore their contribution to the scalar and
pseudoscalar 
matrix elements can arise only from closed loops. It can be shown that 
the contribution of each heavy quark to the scalar matrix element is
given by $<N\vert \bar {Q}_h m_h Q_h \vert N>=\zeta m_N \nbn $ where
$Q_h$ stands for a heavy quark and $\zeta ={2\over 29}$.
 The pseudo-scalar matrix element however receives contribution both from
 light and heavy quark states. The light quark contribution to the 
pseudoscalar matrix element is due to the coupling of $ \bar {u}\imath
\gf u$ and  $ \bar {d}\imath\gf d$
to the nucleon  via the  light intermediate $\pi_0$ state.
They are given by $<N\vert \bar {u}\imath \gf m_u u\vert N>={m_u\over
m_u+m_d}m_N (-g_A)\nbigtn $ and 
 $<N\vert \bar {d}\imath \gf m_d d\vert N>=-{m_d\over
m_u+m_d}m_N (-g_A)\nbigtn $.
The contribution of each heavy quark to the
 pseudoscalar matrix element is given by
$$<N\vert \bar {Q}_h\imath \gf m_h Q_h\vert N>={1\over 2}{m_u-m_d\over
m_u+m_d}m_N(-g_A)\nbigtn-{1\over 2}m_N(-g^0_A)
\nbign .\eqno(5)$$
 For the $Q={2\over 3}$ quarks the net contribution to the pseudoscalar
matrix element is given by [5]
 $$\eqalignno{<N\vert \bar {u}\imath \gf m_u u\vert N>+
 2<N\vert \bar {Q}_h\imath \gf  Q_h\vert N>&={2m_u-m_d\over m_u+m_d}
 m_N(-g_A)\nbigtn \cr
 &-m_N(-g^0_A)\nbign .&(6a)\cr}$$
 
Whereas for the $Q=-{1\over 3}$ quarks  the net contribution is given by [5]
 $$\eqalignno{<N\vert \bar {d}\imath \gf  m_d d\vert N>+
 2<N\vert \bar {Q}_h\imath \gf Q_h\vert N>&={m_u-2m_d\over m_u+m_d}
 m_N(-g_A)\nbigtn \cr
 &-m_N(-g^0_A)\nbign .&(6b)\cr}$$
Here $g_A$ and $g^0_A$ are the axial vector coupling  constants associated
with $<N\vert\partial_{\mu}J^{\mu}_{5,3}\vert N>$ and 
 $<N\vert\partial_{\mu}J^{\mu}_5\vert N>$ respectively. We then get
$$\eqalignno{<N\vert L^{UU}_{PTV}\vert N>&=\zeta m^2_N(-g^0_A)
{\absvo \imath \vPtPt\over \absvt ^2 v}\nbn\nbign\cr
&-\zeta m^2_N(-g_A){2m_u-m_d\over m_u+m_d}
{\absvo \imath \vPtPt\over \absvt ^2 v}
\nbn\nbigtn .&(7 )\cr}$$

Similarly the P and T odd NN interactions arising from the exchange of 
neutral higgs boson between a pair of D quarks and an U and a D quark are
given by 

$$\eqalignno{<N\vert L^{DD}_{PTV}\vert N>&=\zeta m^2_N(-g^0_A)
{\absvt \imath \vPoPt\over \absvo ^2 v}\nbn \nbign\cr
&-\zeta m^2_N(-g_A){m_u-2m_d\over m_u+m_d}
{\absvt \imath \vPoPt\over \absvo ^2 v}
\nbn\nbigtn .&(8 )\cr}$$

and

$$\eqalignno{<N\vert L^{UD}_{PTV}\vert N>&=\zeta m^2_N(-g^0_A)
[{\imath \vPtPt\over \absvo v}+{\imath\vPoPt\over \absvt v}]
\nbn\nbign\cr
&-\zeta m^2_N(-g_A)[{\imath\vPtPt\over \absvo v}{m_u-2m_d\over m_u+m_d}
+{\imath \vPoPt\over \absvt v}{2m_u-m_d\over m_u+m_d}]\times\cr
&\nbn\nbigtn .&(9 )\cr}$$

For $q^2<< m^2_h$ the propagator factors appearing above can be written
as $\vPoPt \approx -\imath \sum_n {C_{1n}C_{3n}\over m^2_n}$ and 
 $\vPtPt \approx -\imath \sum_n {C_{2n}C_{3n}\over m^2_n}$.
 Here the index n stands
for the neutral boson mass eigenstates. The orthogonal $3\times 3$
matrix $C_{in}$
connects the gauge eigenstates $\Phi_i$ with the mass eigenstates
$\tilde {\Phi}_n$. We shall assume that only the lightest neutral higgs (h)
contributes dominantly to the propgators. In that case we get

$$\eqalignno{L^{NN}_{PTV}&=[{C_{1h}C_{3h}v^3\over m^2_h\absvt\absvo^2}+
{C_{2h}C_{3h}v^3\over m^2_H\absvo\absvt^2}]\sqrt {2}G_F\zeta m^2_N
(-g^0_A)\nbn\nbign\cr
&-[{C_{1h}C_{3h}v\over (m_u+m_d)m^2_h}({1\over\absvt} (2m_u-m_d)+
{\absvt\over \absvo ^2}(m_u-2m_d))\cr
&+{C_{2h}C_{3h}v\over (m_u+m_d)m^2_h}({1\over\absvo} (m_u-2m_d)+
{\absvo\over \absvt ^2}(2m_u-m_d))]\sqrt{2}G_F\zeta m^2_N(-g_A)\times\cr
&\nbn\nbigtn .&(10 )\cr}$$

Phenomenologically one can write
 $$L^{NN}_{PTV}\equiv -C^0_{S-PS}{G_F\over \sqrt {2}}\nbn\nbign
 -C^1_{S_PS}{G_F\over \sqrt {2}}\nbn\nbigtn .\eqno(11)$$
where $C^0_{S-PS}$ and $C^1_{S-PS}$ are the parameters
characterizing the strenghts of P and T odd NN interactions [6].
 For the two higgs doublet model under consideration they are given by
$$C^0_{S-PS}=-2\zeta (-g^0_A){m^2_N v^3\over m^2_h\absvo\absvt}
[{C_{1h}C_{3h}\over \absvo}+{C_{2h}C_{3h}\over \absvt}].\eqno(12 )$$
 and
$$\eqalignno{C^1_{S-PS}&=2\zeta (-g_A){m^2_N v\over m^2_h}
[{C_{1h}C_{3h}\over (m_u+m_d)}({1\over\absvt} (2m_u-m_d)+
{\absvt\over \absvo ^2}(m_u-2m_d))\cr
&+{C_{2h}C_{3h}\over (m_u+m_d)}({1\over\absvo} (m_u-2m_d)+
{\absvo\over \absvt ^2}(2m_u-m_d))].&(13)\cr}$$

Note that the isovector coupling constant $C^1_{S-PS}$ vanishes if we set 
$\absvo =\absvt$ and $m_u=m_d$ as required by isospin symmetry. We now
 turn to the values of the parametrs that appear on the rhs of 
the phenomenological constants $C^0_{S-PS}$
and $C^1_{S-PS}$. For an orthogonal matrix $\vert C_{1h}C_{3h}\vert \le .5$
and $\vert C_{2h}C_{3h}\vert \le .5$ . Chiral symmerty breaking 
exhibited in the light
pseudoscalar mass spectrum implies that $m_u=5$ Mev and $m_d=7$ Mev. The 
direct search for the neutral higgs boson by LEP1 has produced the lower bound
$m_h>58.4$ Gev [7] for the three generation SM with minimal Higgs sector
and $m_{h_1}>44$ Gev $(m_{h_1}<m_{h_2})$ for the minimal supersymmetric
SM.
 The CDF group associated with the Fermilab tevatron
has reported having observed the top quark with a mass of around 175 Gev [8].
The CDF result implies a large mass heirarchy between the top quark and 
the bottom quark which can naturally arise from a large value
of ${\absvt\over \absvo}$ if $g_t(m_t)\approx g_b(m_b)$. In the later case
 we get $\absvo\approx 5$ Gev and $\absvt\approx 175 $ Gev. The axial vector 
coupling constants $g^0_A$ and $g_A$ can be assumed to be approximately
equal, i.e. $g^0_A\approx g_A\approx -1.24$. For the above values of the
parameters we find that $\vert C^0_{S-PS}\vert \le 3.08\times 10^{-2}$
and $\vert C^1_{S-PS}\vert \le 2.30\times 10^{-2}$.

 The question that arises
is how does the predicted values of the parameters
compare with the limits derived from
 experiments searching for P and T violating effects in atomic and 
molecular systems. The  P and T odd NN interaction given by eqn. [11] can
 give rise to various T odd nuclear multipoles [6], 
the simplest of which are the
electric dipole moment (EDM), Schiff moment (SM) and magnetic quadrupole
 moment (MQM). The EDM of the nucleus does not lead to an overall
atomic or molecular EDM because of the well known Schiff's theorem [9]. 
Further for
 atoms in the ground state with closed electronic shells the MQM does not
 induce an atomic EDM. So the nuclear SM is the only T odd nuclear multipole 
that leads to an atomic EDM. The SM's of $^{129, 131}$Xe, 
$^{133}$ Cs and $^{199, 201}$ Hg nuclei has been calculated [6] for the
 P and T odd NN interaction given by eqn. [11]. The EDM's of
 Xe, Cs and Hg atoms have also been
 related [6] to their respective SM's using the relativistic  Hartree-Fock
technique. The theoretical calculations when combined with the measurement
 of linear Stark effect in the ground state of Xe, Cs and Hg atoms 
produces the
 following bounds on the isoscalar parameter
 $C^0_{S-PS}$ [6]: $C^0_{S-PS}(Xe)= -.07 \pm .24$,
 $C^0_{S-PS}(Cs)= .6 \pm 2.4 \pm 0.6 $ and $C^0_{S-PS}(Hg)= .01 \pm .03 $.
The expeimental limit obtained from Hg is the most stringent. The reason being
due to larger Z one gains an order of magnitude in the EDM by going from
Xe or Cs to Hg. Experiments with TlF molecule also lead to a bound
on $\cz$, but it is only an order of magnitude estimate because of
theoretical uncertainties involved in estimating the SM of Tl nucleus.
 We find that the predicted values of $C^0_{S-PS}$ in our
 model lie surprisingly close to the experimental limits derived from 
measurements with Hg for which the statistical error is smallest. It is 
therefore extremely important to improve the accuracy for measuring the EDM
of atoms and molecules in order to discriminate the different models of T 
violation on the basis of their closeness to the experimental bound
 and obtain a better physical understanding of its origin.

In conclusion in this article we have estimated the magnitude of the 
parameters that characterize the strength of P and T odd NN interaction 
in the context of a THDM. The model satisfies the
 conditions of tree level NFC. We find that if the ratio of vev's 
${\absvt\over \absvo}\approx {m_t\over m_b}$ then the predicted values of the
parameters are close to the best experimental limits obtained
 from measurements with Hg. A large value of $\tan \beta ={\absvt\over \absvo}$
might naturally explain the large t-b mass splitting. However the same could
 also have other phenomenological effects which should be carefully analysed
before accepting such a scenario. Present experiments however do not rule out
this possibility and that makes the values of $C^0_{S-PS}$ and
$C^1_{S-PS}$ presented in this article somewhat interesting.

\centerline{\bf References}

\item{1.} For a review of SM and its phenomenology see, T. P. Cheng
and L. F. Li, Gauge Theory of Elementary Particle Physice, (OUP, New
York 1984).

\item{2.} H. P. Niles, Phys. Rep. 110, 1 (1984); M. F. Sohinus, ibid. 128, 
2 (1985).

\item{3.} J. F. Donoghue and L. F. Li, Phys. Rev. D 19, 945 (1979);
L. Hall and M. Wise, Nucl. Phys. B 187, 397 (1981).

\item{4.} S. Weinberg, Phys. Rev. D 42, 860 (1990).

\item{5.} M. A. Shifman, A. I. Vainshtein and V. I. Zakharov,
Phys. Lett. B 78, 443 (1978).

\item {6.} V. A. Dzuba, V. V. Flambaum and P. G. Silvestrov, Phys. Lett.
B, 154, 93 (1985); V. V. Flambaum, I. B. Khriplovich and O. P. Sushkov,
ibid. 162, 213 (1985); I. B. Khriplovich, Parity non-conservation
in atomic phenomena (Nauka, Moscow 1981).

\item {7.} R. M. Barnett et al. (Particle data group), Phys. Rev. D 54, 1,
1996.

\item {8.} F. Abe et al. (CDF collaboration), Phys. Rev. Lett. 74, 2626 
(1995)

\item {9.} L. I. Schiff, Phys. Rev. 142, 2194 (1983).

\end